\documentclass[a4paper]{jpconf}

\usepackage{aas_macros}
\usepackage{amssymb,amsmath}
\usepackage{graphicx}
\usepackage[usenames]{xcolor}
\usepackage[normalem]{ulem}
\newcommand{\stkout}[1]{\ifmmode\text{\sout{\ensuremath{#1}}}\else\sout{#1}\fi}

\begin{document}
\title{Magnetic field amplification by the r-mode instability}

\author{Andrey I. Chugunov${}^1$, John L. Friedman${}^2$,  Lee Lindblom${}^{3}$, and
Luciano Rezzolla${}^{4,5}$}

\address{${}^1$Ioffe Institute, Polytekhnicheskaya 26, 194021 St.-Petersburg, Russia}
\ead{andr.astro@mail.ioffe.ru}

\address{${}^2$Leonard Parker Center for Gravitation, Cosmology and Astrophysics,
University of Wisconsin-Milwaukee, P.O. Box 413,
Milwaukee, Wisconsin 53201, USA}
\ead{friedman@uwm.edu}

\address{${}^3$Center for Astrophysics and Space Sciences, University of California at San Diego, La Jolla, CA 92093, USA}
\ead{llindblom@ucsd.edu}

\address{${}^4$Institute for Theoretical Physics,
Max-von-Laue-Stra{\ss}e 1, 60438 Frankfurt, Germany}
\ead{rezzolla@itp.uni-frankfurt.de}

\address{${}^5$Frankfurt Institute for Advanced Studies, Ruth-Moufang-Str. 1,
60438 Frankfurt, Germany}

\begin{abstract}
We discuss magnetic field enhancement by unstable r-modes
(driven by the gravitational radiation reaction force) in
rotating stars. In the absence of a magnetic field,
gravitational radiation exponentially increases the r-mode
amplitude $\alpha$, and accelerates differential rotation
(secular motion of fluid elements). For a magnetized star,
differential rotation enhances the magnetic field energy.
Rezzolla et al.\ (2000--2001) argued that if the magnetic
energy grows faster than the gravitational radiation
reaction force pumps energy into the r-modes, then the
r-mode instability is suppressed.  Chugunov (2015)
demonstrated that without gravitational radiation,
differential rotation can be treated as a degree of freedom
decoupled from the r-modes and controlled by the back
reaction of the magnetic field. In particular, the magnetic
field windup does not damp r-modes. Here we discuss the
effect of the back reaction of the magnetic field on
differential rotation of unstable r-modes, and show that it
limits the generated magnetic field and the magnetic energy
growth rate preventing suppression of the r-mode
instability by magnetic windup at low saturation
amplitudes, $\alpha \ll 1$, predicted by current models.
\end{abstract}

\section{Introduction}

Neutron stars contain matter which is denser than that in
atomic nuclei. Because cold matter at such densities cannot
be produced in terrestrial laboratories, neutron stars
provide a unique opportunity to study the properties of
super-dense matter by comparing observations of these stars
with theoretical models.  One such possibility is
associated with the gravitational radiation driven
instability of r-modes in rapidly rotating neutron stars
\cite{chandrasekhar70a,fs78a,fs78b,Friedman78,Chugunov17,andersson98,fm98}.
In the absence of dissipation and magnetic fields, the
r-modes are unstable in any (even slowly) rotating star
\cite{andersson98,fm98}. Dissipation (e.g., shear and bulk
viscosities \cite{lom98,lmo99}, an Ekman layer
\cite{lu01,ga06a}, mutual friction
\cite{lm00,agh09,hap09,gck14a,gck14b}) can suppress this
instability to some extent. However, if the spin freqeuency
of a neutron star exceeds a (temperature dependent)
critical value, it can be unstable (see e.g.,
\cite{haskell15} for recent review), and the instability
will alter the star's evolution
\cite{olcsva98,levin99,cgk14,Chugunov17}. Observations of
rapidly rotating neutron stars (i.e., neutron stars in
accreting low mass X-ray binaries and their descendants --
millisecond pulsars) put important constraints on the
r-mode instability threshold and thus on properties of
super-dense matter (c.f.\
\cite{hah11,hdh12,as14_msp,Schwenzer_etal_Xray,cgk17,47Tuc_aa}).
However, these constraints assume that the internal fluid
dissipation is the main damping mechanism for the r-modes;
such constraints can be irrelevant if the r-mode
instability is, in fact, suppressed by an alternate
mechanism like a magnetic field windup
\cite{Rezzolla_etal00,Rezzolla_etal01a,Rezzolla_etal01b}.
These publications argue that the growth of unstable
r-modes should be accompanied by differential rotation and
associated bending of the magnetic field lines powered by
the r-mode energy. If the r-mode energy loss exceeds the
energy supplied by the gravitational radiation instability,
the r-modes should be damped out. Here we summarize our
analysis of the back reaction of bent magnetic lines on the
differential rotation, and obtain an upper limit on the the
magnetic field enhancement by the r-mode instability. We
show that magnetic windup cannot suppress the r-mode
instability if the nonlinear saturation amplitude is low,
as predicted by current theoretical models, e.g.\
\cite{btw07,btw09,hga14}. More details can be found in
\cite{flrc17}.

\section{R-modes and differential rotation}

Differential rotation which suppresses the gravitational radiation
driven instability in the magnetic windup scenario
\cite{Rezzolla_etal00,Rezzolla_etal01a,Rezzolla_etal01b}, is a
second-order effect in the r-mode amplitude. Before considering the
full problem (including the effects of gravitational radiation and
magnetic fields), we outline previously published results on two
simplified problems: (1) the r-modes in non-magnetized stars and (2)
stable r-modes (i.e.\ neglecting gravitational radiation) in
magnetized stars.

The solution for stable r-modes in non-magnetized stars to
second-order in amplitude $\alpha$ was obtained in
\cite{Sa04}. The axisymmetric part of this solution
describes differential rotation and contains a gauge
dependent part; the general solution includes contribution,
corresponding to the second order differential rotation
allowed in non-oscilating star (i.e., an arbitrary time
independent azimuthal flow, stratified on cylinders). A
particular solution  with vanishing drift velocity (no
secular motion of fluid elements) exists. However, the
latter solution does not mean  that the axisymmetric
oscillation averaged second order velocity perturbations
are absent. On the contrary, according to \cite{Sa04} these
terms are not zero and are not cylindrically stratified for
any choice of initial conditions. The apparent
contradiction between the absence of secular motion of the
fluid elements and non-vanishing velocity perturbations is
superficial. It is associated with Stokes drift
\cite{Stokes1847,LH53} -- second order secular motion of
fluid elements associated with the oscillating velocity
profile of the first-order perturbation. The Stokes drift
contributes to the secular motion along with the
oscillation averaged velocity perturbations and resolves
the contradiction (see \cite{Chugunov15} for details).

Recently,  an analytical second order solution for the unstable
r-modes in non-magnetized rotating stars was obtained  \cite{fll16},
neglecting internal fluid dissipation.  This solution demonstrates
an exponential growth of the r-mode amplitude,
$\alpha(t)=\alpha(0)\exp(t/\tau_\mathrm{GR})$ (with the time scale
$\tau_\mathrm{GR}>0$ determined previously in \cite{lom98}). The
growth is accompanied by the exponential growth of differential
rotation (with a particular spatial profile determined by the
equation of state). Like a stable r-mode, this solution has a gauge
freedom -- the general solution includes an arbitrary second order
time independent differential rotation profile. The secular motion
of fluid elements is still cylindrically stratified, but the
solution with vanishing drift is absent; the exponential growth of
velocity perturbations leads to an exponential growth of fluid
element displacements.

For a magnetized star with perfect conductivity, the
secular motion of fluid elements leads to bending of
magnetic field lines. A solution with time independent
non-vanishing drift velocity becomes impossible. This was
confirmed in \cite{Chugunov15} where the general second
order solution for the stable r-modes in a magnetized
neutron star was obtained. As with non-magnetized stars,
the general solution has gauge freedom; it is associated
with perturbations of a non-oscillating star and  can be
described as an ensemble of Alfv\'en modes.  A solution
with  vanishing drift of fluid elements exists for specific
initial conditions.  The general second order solution can
be presented as a superposition of two solutions: (a) a
solution which describes the evolution of differential
rotation in a non-oscillating magnetized star (i.e. an
ensemble of Alfv\'en modes, determined by initial
conditions) and (b) the r-mode solution with vanishing
drift.  This solution demonstrates that stable r-modes are
not damped by magnetic windup.

The criteria for the gravitation radiation driven instability  was generalized to magnetized stars in \cite{ga07}. However
this result cannot be directly applied to check the magnetic windup
scenario.  This scenario assumes the r-mode instability is not
suppressed initially, but it can be suppressed by the growth of the
unstable r-mode leading to windup of the magnetic field and
subsequent suppression of the instability.

The magnetic field evolution in the presence of an unstable
r-mode is analyzed in our recent paper \cite{flrc17} using
second order perturbation theory for an infinitely
conducting rotating star including gravitational radiation
reaction forces. The results are illustrated there by a
simple toy model, and confirmed by a detailed analysis of
the full problem using the symplectic product formalism
\cite{fs78a,ds79}. They can be summarized as follows. In
second order perturbation theory, the gravitational
radiation reaction and magnetic forces have axisymmetric
components with non-vanishing oscillation averages that
directly affect the secular motion of fluid elements. These
components are (a) a gravitational radiation reaction
force, with magnitude (per unit mass) $f_\mathrm{GR}\sim
\alpha^2(t)\, \tau_\mathrm{GR}^{-1}\, \Omega R$, where
$\Omega=2\pi\nu$ is the spin frequency of the star, and (b)
an effective  force associated with the second order terms
in the perturbation equations. These terms have a part
independent of the magnetic field with magnitude of order
$f_{GR}$ and a part from magnetic terms with magnitude (per
unit mass) $f_\mathrm{m}\sim \alpha^2(t)\, \omega_\mathrm A
\Omega R$ \cite{flrc17}, assuming a non-superconducting
stellar interior. Here, $\omega_\mathrm A=\sqrt{\pi
B^2/\rho R^2}$ is a typical frequency of Alfv\'en modes,
$B$ is an initial magnetic field, $\rho$ is a typical
density, and $R$ is the stellar radius. Before saturation,
when the mode amplitude $\alpha(t)$ is exponentially
growing, $\alpha(t)=\alpha(0)\exp(t/\tau_\mathrm{GR})$, the
displacements of fluid elements in the azimuthal direction
can be estimated in the same way as the displacement of an
``Alfv\'en'' harmonic oscillator with frequency
$\omega_\mathrm A$ evolving under the action of an
exponentially growing external force
$f_\mathrm{GR}+f_\mathrm{m}$ (with timescale
$\tau_\mathrm{GR}/2$). The result is \cite{flrc17}
\begin{equation}
\xi^{\hat\phi}\sim \frac{f_\mathrm{GR}+f_\mathrm{m}}{\left(2\tau_\mathrm{GR}^{-1}\right)^2+\omega_\mathrm A^2}\sim
 \alpha^2(t)\, \Omega  R\,
\frac{\mathrm{max}(\omega_\mathrm A,\,
\tau_\mathrm{GR}^{-1})}{4\tau_\mathrm{GR}^{-2}+\omega_\mathrm
A^2} \le \alpha^2(t)\,R\,
\frac{\Omega}{\omega_\mathrm A}, \label{xi_exp}
\end{equation}
where the last inequality follows from a detailed analysis
of the terms. This estimate also allows us to estimate the
magnetic field energy density,
\begin{equation}
{\epsilon_\mathrm{m}}\sim \rho\, \omega_\mathrm{A}^2\left(\xi^{\hat\phi}\right)^2
\le  \alpha^4(t)\, \rho\,\Omega^2 R^2 \label{em}
\end{equation}
and its growth rate
$\dot \epsilon_\mathrm{m}\sim
4\alpha^4(t)\,\tau_\mathrm{GR}^{-1}\, \rho\, \Omega^2 R^2$,  
again using the harmonic oscillator analogy, or a more detailed
mathematical analysis of the full problem \cite{flrc17}.
The energy density of r-modes can be estimated as
$\epsilon_\mathrm{r}\sim \alpha^2(t)\,\rho\Omega^2 R^2$ (c.f.\
\cite{lom98}), and its  rate as $\dot \epsilon_\mathrm{r}\sim
2\alpha^2(t)\, \tau_\mathrm{GR}^{-1}\,\rho\, \Omega^2 R^2\sim
\alpha^{-2}(t) \dot \epsilon_\mathrm{m}\gg \dot
\epsilon_\mathrm{m}$. Therefore, the magnetic wind up cannot
suppress the r-mode instability before saturation (unless the
saturation amplitude $\alpha(t)\sim 1$, as assumed in early papers
\cite{Rezzolla_etal00,Rezzolla_etal01a,Rezzolla_etal01b}).

In the simplest model of nonlinear r-mode saturation, the amplitude
growth simply stops and remains frozen once
$\alpha(t)=\alpha_\mathrm{sat}$.  In this case, $\xi^{\hat \phi}$
can be estimated as the displacement of a harmonic Alfv\'en
oscillator under the influence of the constant force
$f_\mathrm{GR}+f_\mathrm{m}$ with mode amplitude
$\alpha=\alpha_\mathrm{sat}$,
\begin{equation}
\xi^{\hat\phi}_\mathrm{sat}\sim
 \frac{f_\mathrm{GR}+f_\mathrm{m}}{\omega_\mathrm A^2}
 \sim
\alpha^2_\mathrm{sat}\, \Omega  R\,
\frac{\mathrm{max}(\omega_\mathrm A,\,
\tau_\mathrm{GR}^{-1})}{\omega_\mathrm A^2}. \label{xi_sat}
\end{equation}
It can significantly exceed the displacement in the
exponential growth phase, $\xi^{\hat\phi}$, if
$\omega_\mathrm A\ll \tau_\mathrm{GR}^{-1}$.  The growth
rate of the magnetic energy can be estimated as
 $\dot
\epsilon_\mathrm{m}\sim
\omega_\mathrm{A}\epsilon_\mathrm{m}$, with
$\epsilon_\mathrm{m}$ given by (\ref{em}) with
$\xi^{\hat\phi}=\xi^{\hat\phi}_\mathrm{sat}$.  Formally,
$\dot \epsilon_\mathrm{m}$ is comparable to $\dot
\epsilon_\mathrm{r}$ if $\omega_\mathrm{A}\sim
\alpha_\mathrm{sat}^2 \tau_\mathrm{GR}^{-1}\ll
\tau_\mathrm{GR}^{-1}$. However, even for an unexpectedly
large $ \alpha_\mathrm{sat}\sim 10^{-3}$ in a
non-superconducting neutron star, this will affect the
r-mode instability only if the initial magnetic field is
$\lesssim 100$\,G. Otherwise the magnetic windup cannot
suppress the r-mode instability.  Equation (\ref{xi_sat})
can also be used to estimate the maximum azimuthal magnetic
field, generated by a magnetic windup in a
non-superconducting star,
\begin{equation}
\delta B^{\hat{\phi}}\sim B\,\frac{\xi^{\hat{\phi}}}{R}\lesssim
 4\times 10^8\
 \left(\frac{\alpha}{10^{-4}}\right)^2
 \frac{\mathrm{max}(\omega_\mathrm A,\, \tau_\mathrm{GR}^{-1})}{\omega_\mathrm A}\,
 \frac{\nu}{500\,\mathrm{Hz}}\,
 \frac{R}{10^6\, \mathrm{cm}}\,
  \frac{\rho}{4\times 10^{14}\,\mathrm{g\,cm}^{-4}}\,
  \mathrm{ G}.
\end{equation}

\section{Summary and conclusions}

We have outlined nonlinear (second order in perturbation
theory) unstable r-modes in magnetized stars composed of a
perfect fluid with infinite conductivity.  Previously, the
second order r-modes were studied for non-magnetized and
magnetized Newtonian stars (\cite{Sa04} and
\cite{Chugunov15}, respectively), and for non-magnetized
stars with gravitational radiation reaction \cite{fll16}.
Our approach here includes both gravitational radiation
reaction and magnetic forces and, therefore, provides a
self-consistent description of the r-mode instability in
magnetized stars. This problem is important because of the
magnetic windup scenario
\cite{Rezzolla_etal00,Rezzolla_etal01a,Rezzolla_etal01b} in
which the r-mode instability leads to magnetic field
enhancement that suppresses the r-mode instability.  Our
results confirm the suggestion
\cite{Rezzolla_etal00,Rezzolla_etal01a,Rezzolla_etal01b},
that the growing r-modes induce the differential drift of
fluid elements and generation of magnetic fields.  However,
before nonlinear saturation is achieved [i.e., at the stage
of exponentially growing $\alpha(t)$], the magnetic energy
growth rate is a factor of $\alpha^2(t)$ smaller than the
r-mode energy growth rate by radiation reaction.  Thus
magnetic windup cannot affect the r-mode instability if
$\alpha(t)\ll 1$. After the saturation, the enhancement of
the magnetic energy is also strongly restricted. Current
models of nonlinear saturation \cite{btw07,btw09,hga14}
suggest that the r-mode amplitude is limited by $\alpha\sim
\alpha_\mathrm{sat}\lesssim 10^{-4}$. Therefore, the
magnetic windup cannot suppress the instability in this
case. We have also estimated the maximal magnetic field
that could be generated by the r-mode instability. For an
unexpectedly large saturation amplitude
$\alpha_\mathrm{sat}=10^{-3}$, an initial magnetic field
$10^8$\,G could be amplified up to about $10^{11}$~G, which
can be important for the magnetic field generation in
neutron stars. However if the initial magnetic field is
$\sim 10^{10}$~G, it would not be significantly affected by
the r-mode magnetic windup.  Our analysis assumes a
background star with no unstable or marginally stable
axisymmetric modes.  Once differential rotation is
established, however, a magnetorational instability (MRI)
is likely.  MRI-unstable perturbations cannot acquire more
energy than is present in the small available differential
rotation, and we therefore suspect that the presence of
MRI-unstable or marginally unstable perturbations will not
substantially alter our analysis. But this presumption
should be carefully checked in the future.

\ack

 We thank an anonymous referee for useful suggestions that improved this
  paper and Ruxandra Bondarescu, Mikhail E.\ Gusakov, Branson Stephens,
 and Ira Wasserman for helpful conversations.  AIC was supported by
 the Russian Science Foundation (Grant No. 14-12-00316). LL was
 supported in part by NSF grants PHY 1604244 and DMS 1620366 to the
 University of California at San Diego. LR was supported in part by
 ''NewCompStar'', COST Action MP1304, from the LOEWE-Program in HIC
 for FAIR, the European Union's Horizon 2020 Research and Innovation
 Programme under grant agreement No. 671698 (call FETHPC-1-2014,
 project ExaHyPE), from the ERC Synergy Grant ``BlackHoleCam - Imaging
 the Event Horizon of Black Holes'' (Grant 610058), and from JSPS
 Grant-in-Aid for Scientific Research(C) No. 26400274.

\section*{References}
\providecommand{\newblock}{}


\providecommand{\newblock}{}
\begin{thebibliography}{}
\expandafter\ifx\csname url\endcsname\relax
  \def\url#1{{\tt #1}}\fi
\expandafter\ifx\csname urlprefix\endcsname\relax\def\urlprefix{URL }\fi
\providecommand{\eprint}[2][]{\url{#2}}

\end{thebibliography}


\begin{thebibliography}{10}
    \expandafter\ifx\csname url\endcsname\relax
    \def\url#1{{\tt #1}}\fi
    \expandafter\ifx\csname urlprefix\endcsname\relax\def\urlprefix{URL }\fi
    \providecommand{\eprint}[2][]{\url{#2}}

    \bibitem{chandrasekhar70a}
    {Chandrasekhar} S 1970 {\em \prl\/} {\bf 24} 611--615

    \bibitem{fs78a}
    {Friedman} J~L and {Schutz} B~F 1978 {\em \apj\/} {\bf 221} 937--957

    \bibitem{fs78b}
    {Friedman} J~L and {Schutz} B~F 1978 {\em \apj\/} {\bf 222} 281--296

    \bibitem{Friedman78}
    {Friedman} J~L 1978 {\em Communications in Mathematical Physics\/} {\bf 62}
    247--278

    \bibitem{Chugunov17}
    {Chugunov} A~I 2017 {\em \pasa\/} {\bf 34} e046

    \bibitem{andersson98}
    {Andersson} N 1998 {\em \apj\/} {\bf 502} 708

    \bibitem{fm98}
    {Friedman} J~L and {Morsink} S~M 1998 {\em \apj\/} {\bf 502} 714

    \bibitem{lom98}
    {Lindblom} L, {Owen} B~J and {Morsink} S~M 1998 {\em \prl\/} {\bf 80}
    4843--4846 

    \bibitem{lmo99}
    {Lindblom} L, {Mendell} G and {Owen} B~J 1999 {\em \prd\/} {\bf 60} 064006

    \bibitem{lu01}
    {Levin} Y and {Ushomirsky} G 2001 {\em \mnras\/} {\bf 324} 917--922

    \bibitem{ga06a}
    {Glampedakis} K and {Andersson} N 2006 {\em \prd\/} {\bf 74} 044040

    \bibitem{lm00}
    {Lindblom} L and {Mendell} G 2000 {\em \prd\/} {\bf 61} 104003

    \bibitem{agh09}
    {Andersson} N, {Glampedakis} K and {Haskell} B 2009 {\em \prd\/} {\bf 79}
    103009 

    \bibitem{hap09}
    {Haskell} B, {Andersson} N and {Passamonti} A 2009 {\em \mnras\/} {\bf 397}
    1464--1485 

    \bibitem{gck14a}
    {Gusakov} M~E, {Chugunov} A~I and {Kantor} E~M 2014 {\em \prl\/} {\bf 112}
    151101 

    \bibitem{gck14b}
    {Gusakov} M~E, {Chugunov} A~I and {Kantor} E~M 2014 {\em \prd\/} {\bf 90}
    063001 

    \bibitem{haskell15}
    {Haskell} B 2015 {\em International Journal of Modern Physics E\/} {\bf 24}
    1541007 

    \bibitem{olcsva98}
    {Owen} B~J, {Lindblom} L, {Cutler} C, {Schutz} B~F, {Vecchio} A and {Andersson}
    N 1998 {\em \prd\/} {\bf 58} 084020

    \bibitem{levin99}
    {Levin} Y 1999 {\em \apj\/} {\bf 517} 328--333 

    \bibitem{cgk14}
    {Chugunov} A~I, {Gusakov} M~E and {Kantor} E~M 2014 {\em \mnras\/} {\bf 445}
    385--391 

    \bibitem{hah11}
    {Ho} W~C~G, {Andersson} N and {Haskell} B 2011 {\em \prl\/} {\bf 107} 101101

    \bibitem{hdh12}
    {Haskell} B, {Degenaar} N and {Ho} W~C~G 2012 {\em \mnras\/} {\bf 424} 93--103

    \bibitem{as14_msp}
    {Alford} M~G and {Schwenzer} K 2014 {\em Physical Review Letters\/} {\bf 113}
    251102 

    \bibitem{Schwenzer_etal_Xray}
    {Schwenzer} K, {Boztepe} T, {G{\"u}ver} T and {Vurgun} E 2017 {\em \mnras\/}
    {\bf 466} 2560--2569 

    \bibitem{cgk17}
    {Chugunov} A~I, {Gusakov} M~E and {Kantor} E~M 2017 {\em \mnras\/} {\bf 468}
    291--304 

    \bibitem{47Tuc_aa}
    {Bhattacharya} S, {Heinke} C~O, {Chugunov} A~I, {Freire} P~C~C, {Ridolfi} A and
    {Bogdanov} S 2017
    {\em \mnras\/}
    {\bf 472} 3706--3721

    \bibitem{Rezzolla_etal00}
    {Rezzolla} L, {Lamb} F~K and {Shapiro} S~L 2000 {\em \apjl\/} {\bf 531}
    L139--L142 

    \bibitem{Rezzolla_etal01a}
    {Rezzolla} L, {Lamb} F~K, {Markovi{\'c}} D and {Shapiro} S~L 2001 {\em \prd\/}
    {\bf 64} 104013 

    \bibitem{Rezzolla_etal01b}
    {Rezzolla} L, {Lamb} F~K, {Markovi{\'c}} D and {Shapiro} S~L 2001 {\em \prd\/}
    {\bf 64} 104014 

    \bibitem{btw07}
    {Bondarescu} R, {Teukolsky} S~A and {Wasserman} I 2007 {\em \prd\/} {\bf 76}
    064019 

    \bibitem{btw09}
    {Bondarescu} R, {Teukolsky} S~A and {Wasserman} I 2009 {\em \prd\/} {\bf 79}
    104003 

    \bibitem{hga14}
    {Haskell} B, {Glampedakis} K and {Andersson} N 2014 {\em \mnras\/} {\bf 441}
    1662--1668 

    \bibitem{flrc17}
    {Friedman} J~L, {Lindblom} L, {Rezzolla} L and {Chugunov} A~I 2017
    {\em \prd\/} {\bf 96} 124008 

    \bibitem{Sa04}
    {S{\'a}} P~M 2004 {\em \prd\/} {\bf 69} 084001

    
    \bibitem{Stokes1847}
    Stokes G~G 1847 {\em Transactions of the Cambridge Philosophical Society\/}
    {\bf 8} 441--455

    \bibitem{LH53}
    {Longuet-Higgins} M~S 1953 {\em Phil. Trans. R. Soc. Long. A\/} {\bf 245} 535

    \bibitem{Chugunov15}
    {Chugunov} A~I 2015 {\em \mnras\/} {\bf 451} 2772--2779
    
    \bibitem{fll16}
    Friedman J~L, Lindblom L and Lockitch K~H 2016 {\em Phys. Rev. D\/} {\bf 93}(2)
    024023

    \bibitem{ga07}
    {Glampedakis} K and {Andersson} N 2007 {\em \mnras\/} {\bf 377} 630--644

    \bibitem{ds79}
    {Dyson} J and {Schutz} B~F 1979 {\em Proceedings of the Royal Society of London
        Series A\/} {\bf 368} 389--410

\end{thebibliography}
\end{document}